\documentclass[twocolumn,showpacs,showkeys,
preprintnumbers,amsmath,amssymb,aps]{revtex4}
\usepackage{graphicx}
\usepackage{dcolumn}
\usepackage{bm}
\usepackage{longtable}
\newcommand{\thmeff}{thermodynamic efficiency}

\begin{document}
\title{Stoke's efficiency of temporally rocked ratchets}
\author{Raishma Krishnan}
\author{Jim Chacko}
\author{Mamata Sahoo}
\author { A. M. Jayannavar }
\email{jayan@iopb.res.in}
\affiliation{Institute of Physics, Sachivalaya Marg, 
Bhubaneswar 751005, India }

\begin{abstract}                          
We study the generalized efficiency of an adiabatically rocked ratchet with 
both spatial and temporal asymmetry. We obtain an analytical 
expression for the generalized efficiency in the 
deterministic case. Generalized efficiency of the
order of $50\%$ is obtained by fine tuning of the parameter range. This is
unlike the case of thermodynamic efficiency where we could readily 
get an enhanced efficiency of upto $90\%$.
The observed higher values of generalized efficiency is attributed to be 
due to the suppression of backward current. We have also discussed briefly 
the differences between thermodynamic, rectification or generalized efficiency 
and Stoke's efficiency. Temperature is found to optimize the 
generalized efficiency over a wide range of parameter space unlike in
 the case of \thmeff.
\vskip.5cm
\end{abstract}


\maketitle

\section{Introduction}

Nonequilibrium fluctuations can induce directed transport along periodic
extended structures without the application of a net external bias. 
Diverse studies exist in literature which 
centralize on this phenomena of noise induced 
transport~\cite{julicher,reiman,1amj,special}. The extraction 
of useful work by the rectification of thermal fluctuations 
inherent in the medium at the expense of an overall increase in the
entropy of the system plus the environment~\cite{parrondo,demon,2amj} 
have become a major area of research in nonequilibrium statistical 
mechanics. The key criterion for the possibility of such a 
transport are the presence of unbiased nonequilibrium perturbations 
and a broken spatial or temporal symmetry. With the increase in prominence of 
the study of nano-size particles, the concurrent thermal agitations can no 
longer be ignored. The perceptivity of the basic mechanism of 
ratchet operation has been disclosed through various models 
like flashing ratchets, rocking ratchets, time asymmetric 
ratchets, frictional ratchets etc~\cite{julicher,reiman,1amj,special}.
 
Extensive studies have been done to understand the nature of currents, 
their possible reversals and also the efficiency of energy transduction. These 
results are of immense utilization in the  development of 
proper models that efficiently separate particles of micro and 
nano sizes and also in turn for the development of machines at nano
scales~\cite{raishma-natl}. Processes in which the chemical energy stored 
in a nonequilibrium bath is transformed into useful work are 
believed to be the basis of molecular motors and are of great importance 
in active biological processes.

With the development of a separate subfield called stochastic 
energetics~\cite{sekimoto,parrondo}, the reaction force 
exerted by the stochastic system on the bath is identified with the 
heat discarded by the system to the bath. With this definition, it has become 
possible to establish the compatibility between the 
Langevin or Fokker-Planck formalism with the laws of 
thermodynamics. This framework helps to calculate 
various physical quantities  like efficiency of energy 
transduction~\cite{kamgawa}, energy dissipation 
(hysteresis loss), entropy production~\cite{rkamj} etc., 
thereby rendering a new tool to study systems far from equilibrium. 
  
In the present work we consider time asymmetric ratchet 
~\cite{pre,chialvo,ai} where the ratchet potential is rocked adiabatically in
time in such a way that a large force field $F(t)$ acts for a 
short time interval of period in the forward direction 
as compared to a smaller force field for a
longer time interval in the opposite direction. The intervals 
are so chosen that the net external force or bias acting on the particle over 
a period is zero. With such a time asymmetric forcing, one can generate 
enhanced unidirectional currents even in the presence of a spatially
symmetric periodic potential~\cite{chialvo}. 

\begin{figure}[hbp!]
\begin{center}
\input{epsf}
\includegraphics [width=2.5in,height=2in]{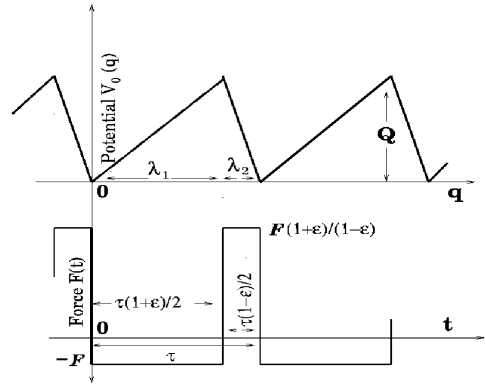}
\caption{Plot of sawtooth potential as a function of coordinate $q$ 
and the time asymmetric forcing $F(t)$ as a function of $t$. }
\label{sawtooth}
\end{center}
\end{figure}

The schematic figure of the ratchet potential $V_0(q)$ chosen for our present 
work and the time asymmetric forcing $F(t)$ are shown in 
Fig.~\ref{sawtooth}. Time asymmetric forcing can also be generated by applying 
a biharmonic force or harmonic mixing ~\cite{marchi}. Theoretically, 
time asymmetric ratchets have been considered in 
earlier literatures under different physical contexts 
~\cite{save,chacon}. Several experimental studies
have also been explored such as generation of photo-currents in semiconductors
 ~\cite{shmelev}, transport in binary mixtures~\cite{save}, 
realization of Brownian motors using 
cold atoms in  a dissipative optical lattice~\cite{schiavoni} etc.

One of the key concepts in the study of the performance characteristics of 
Brownian engines/ratchets is the notion of
efficiency of energy transduction from the fluctuations~\cite{linke}. 
The primary need for efficient motors arises either to 
decrease the energy consumption rate and/or to decrease the heat dissipation in
the process of operations and it is the latter concept which is of
more importance in the present world of miniaturization 
of components~\cite{parrondo}. As the ratchet operates in 
a nonequilibrium state there is always an unavoidable 
and irreversible transfer of heat via fluctuations (in coordinate and
accompanying velocity) thereby making it 
less efficient as a motor. Any irreversibility or finite entropy production 
will reduce the efficiency. For instance, the attained value of thermodynamic 
efficiency in flashing and rocking ratchet are below the subpercentage regime
($<0.01$). However, it has been shown that at very low temperatures 
fine tuning of parameters could easily lead to a larger 
efficiency, the regime of parameters being very narrow~\cite{sokolov}. 
Protocols to optimize the efficiency in saw tooth ratchet potential 
in presence of spatial symmetry and symmetric temporal rocking 
have been worked out in detail in~\cite{sokolov,hernandez}. 

By construction of a special type of flashing ratchet with two 
asymmetric double-well periodic-potential
states displaced by half a period~\cite{tsong} a high efficiency of 
an order of magnitude higher than in earlier models
~\cite{sekimoto,parrondo,kamgawa,astu05} were obtained. The basic essence 
here was that even for diffusive Brownian motion the 
choice of appropriate potential profile ensures suppression of 
backward motion leading to a reduction in the 
accompanying dissipation. Similar to the case of flashing ratchets
~\cite{tsong} we had earlier studied the motion of 
a particle in a rocking ratchet by applying a temporally asymmetric 
but unbiased periodic forcing in 
the presence of a sinusoidal~\cite{pre} and saw tooth potential~\cite{jstat}. 
The efficiency obtained was very high, much above the 
subpercentage level, about $\sim 30-40 \%$, without fine tuning 
for the case of sinusoidal and $\sim 90\%$ for the saw tooth case 
in the presence of temporal asymmetry.

It is to be pointed that in all ratchet models the particles  move 
in a periodic potential system and hence it ends up with the 
same potential energy even after crossing over to the adjacent 
potential minimum. There is no extra energy stored 
in the particle which can be usefully expended when needed. 
Hence to have an engine out of a ratchet it is necessary to use 
its systematic motion to store potential energy which 
inturn is achieved if a ratchet lifts a load ~\cite{parrondo,kamgawa-parrondo}.
Thus a load force $L$ is applied in a direction opposite to the 
direction of current in the ratchet. With this definition, the
thermodynamic efficiency assumes a zero value 
when no load force is acting~\cite{parrondo,machura}. 

However, as not all motors are designed to pull the loads alternate
proposals for efficiency have come up depending on the task 
the motor have been proposed to do without taking recourse to the 
application of a load force. Some motors may have to achieve high velocity 
against a frictional drag. This consecutively implies that the objective
of the motor considered is to move a certain distance in a given 
time interval with minimal fluctuations in velocity and its position. 
In such a case one defines the generalized efficiency~\cite{derenyi} 
or rectification efficiency~\cite{munakata} which in the absence of 
load is sometimes called as {\it{Stokes}} efficiency~\cite{oster}, given
by the expression 
\begin{equation}
\eta_{S}=\frac{E_{min}}{E_{in}}.
\end{equation}
$E_{min}=\gamma<v>^2$ is the minimum average power necessary to 
maintain the motion of the motor with an average velocity $<v>$ 
against an opposing frictional force. $E_{in}$ is the average input power. 
In the presence of load 
the {\it{generalized efficiency}} or {\it{rectification efficiency}} 
is defined as ~\cite{machura1,derenyi,munakata}, $\eta_{g-r}$ 
\begin{equation}
\eta_{g-r}=\frac{L<v>+\gamma<v>^2}{E_{in}}
\end{equation}
The numerator in the above equation is the sum of the average 
power necessary to move against the external load and against 
the frictional drag with velocity $<v>$ respectively. The thermodynamic 
efficiency of energy transduction is given by
\begin{equation}
\eta_{t}=\frac{L<v>}{E_{in}}.
\end{equation}
This definition can be used for the overdamped~\cite{munakata} 
as well as in the underdamped case~\cite{machura}. For
the case of underdamped Brownian motor there is an added advantage that 
the input power can be written in terms of experimentally 
observable quantities namely, $<v>$ and its
fluctuations~\cite{machura,munakata}. 
This is independent of the model of the ratchet chosen. 

{\it{In the present work we mainly 
analyze the nature of generalized efficiency in the absence 
of load}}, namely Stoke's efficiency. The behavior in presence of 
load is also briefly discussed. We obtain values of Stoke's efficiency 
of the order of $50\%$ by fine tuning the parameters. In the 
generic parameter space, we obtain efficiencies much above the 
sub percentage regime. The earlier model for the case of 
flashing ratchet~\cite{munakata} gave a generalized efficiency 
of the order of $0.2$. In a recent study~\cite{machura2}, 
it has been shown that Stoke's 
efficiency exhibits a high value of around $0.75$, when the motor
operates in an inertial regime and at very low temperatures. 
However, these inertial motors do not exhibit high thermodynamic 
efficiency. We also show that unlike thermodynamic 
efficiency, the generalized efficiency  
is aided or optimized by temperature.

Our paper is organized as follows. We first describe our model in
Section \ref{sect2}. Results and  discussions 
are given in Section~\ref{sect3} which is consecutively 
followed by conclusions in Section~\ref{sect4}.

\section{The Model:}\label{sect2}

Our model consists of an overdamped Brownian particle with co-ordinate 
$q(t)$ in a spatially asymmetric potential $V(q)$ subjected 
to a temporally asymmetric rocking. The stochastic 
differential equation or the Langevin equation for such a 
particle is given by~\cite{overdamp}
\begin{equation}
 {\dot{q}} = {- {\frac{(V^\prime(q)-F(t))}{\gamma}}}
 +  \xi(t),\label{lang}
 \end{equation}
with $\xi(t)$ being the randomly fluctuating Gaussian 
thermal noise having zero mean and correlation, 
$<\xi(t)\xi(t^\prime)>\,=\,(2\,{k_BT}/\gamma) \delta(t-t^\prime)$ with 
$\gamma$ being the friction coefficient. We consider in the present work a 
piecewise linear ratchet potential  as in 
the case of Magnasco ~\cite{magnasco} with periodicity
$\lambda=\lambda_1+\lambda_2$ set equal
to unity, Fig.~\ref{sawtooth} . This also corresponds 
to the spacing between the wells. We later on scale all the 
lengths with respect to $\lambda$. 

\begin{eqnarray}
V(q) &=& \frac{Q}{\lambda_1} q, \,\,\,\,\,\,\,\, q\leq \lambda_1 \nonumber \\
       &=& \frac{Q}{\lambda_2} (1-q), \,\, \lambda_1<q\leq \lambda
\end{eqnarray}
$F(t)$ which is the externally applied time asymmetric force with zero average 
over the period is also shown in
Fig.~\ref{sawtooth}.  The force in the gentler and
steeper side of the potential are respectively $f^+=\frac{-Q}{\lambda_1}$  
and $f^-= \frac{Q}{\lambda_2}$ and $Q$ is the height of the potential.  

We are interested in the adiabatic rocking regime where the forcing $F(t)$ 
is assumed to change very slowly, 
i.e., its frequency is smaller than any other 
frequency related to the relaxation rate in the problem 
such that the system is in a steady state at each instant of time. 

Following Stratonovich interpretation~\cite{stratanovich}, 
the corresponding Fokker-Planck equation~\cite{riskin} is given by
\begin{eqnarray}
\frac {\partial P(q,t)}{\partial t}&=& \frac {\partial}{\partial q}
\Big[k_BT \frac{\partial P(q,t)}{\partial q}\\ \nonumber
&+& 
[V^\prime(q)-F(t)+L]P(q,t)\Big].\label{fk}
\end{eqnarray} 

The probability current density $j$ for the case of constant force (or static 
tilt) F is given by
\begin{widetext}
\begin{equation}
j(F_0)= \frac{{P_2}^2
\sinh\{\lambda[F_0-L]/2k_BT\}}{{{k_BT{(\lambda/Q)}^2}P_3}
-
{(\lambda/Q)P_1P_2 \sinh \{\lambda[F_0-L]/2k_BT\}}}
\end{equation}
\end{widetext}
where 
\begin{eqnarray}
P_1 &=& \Delta + \frac{\lambda^2-\Delta^2}{4} \frac{F_0-L}{Q}\\
P_2 &=& (1-\frac{\Delta [F_0-L]}{2Q})^2-(\frac{\lambda[F_0-L]}{2Q})^2\\
P_3 &=& \cosh(\{Q-\Delta/2[F_0-L]\}/k_BT)- \nonumber \\
    &  & \mbox{}  \cosh\{\lambda[F_0-L]/2k_BT\}
\end{eqnarray}
where $\Delta=\lambda_1-\lambda_2$ is the spatial asymmetry factor. 
In the above expression we have also included
the presence of an external load $L$, which is essential for defining
thermodynamic efficiency. The current in the 
stationary adiabatic regime averaged over the period $\tau$ 
of the driving force $F(t)$  is given by 
\begin{equation}
<j>=\frac{1}{\tau}\, \int_0^\tau \, j(F(t))\,dt.
\end{equation}    
The form of the  time asymmetric ratchets 
with a zero mean periodic driving force that we have chosen 
~\cite{pre,chialvo,ai} is given by
\begin{eqnarray}
F(t)&=& \frac{1+\epsilon}{1-\epsilon}\, F,\,\, (n\tau 
\leq t < n\tau+ \frac{1}{2} \tau (1-\epsilon)), \\ \nonumber
    &=& -F,\,\, (n\tau+\frac{1}{2} \tau(1-\epsilon) < t \leq
(n+1)\tau).\label{ft}
\end{eqnarray}
Here, the parameter $\epsilon$ signifies the temporal asymmetry 
in the periodic forcing, $\tau$ the period of the driving force $F(t)$ 
and $n=0,1,2....$ is an integer. For this forcing in the adiabatic limit the 
expression for time averaged current is~\cite{chialvo,kamgawa}
\begin{eqnarray}
<j>=j^+ + j^-\,,
\end{eqnarray}
with 
\begin{eqnarray}
j^+&=& \frac{1}{2}(1-\epsilon)\, j(\frac{1+\epsilon}{1-\epsilon}F)\,,
\\ \nonumber \label{jplus}
j^- &=& \frac{1}{2}(1+\epsilon)\,j(-F) \label{jminus} 
\end{eqnarray}
where $j^+$ is the 
current fraction in the positive direction over a fraction of time period 
$(1-\epsilon)/2$ of $\tau$ when the external driving force field is 
$(\frac{1+\epsilon}{1-\epsilon})F$ and $j^-$ is 
the current fraction over the  time period 
$(1+\epsilon)/2$ of $\tau$ when the external driving force 
field is $-F$. 
The input energy $E_{in}$ per unit time is given by~\cite{kamgawa,pre}
\begin{eqnarray}
E_{in}=F [(\frac{1+\epsilon}{1-\epsilon})j^+-j^-].\label{Ein}
\end{eqnarray}  

In order for the system to do useful work, a load force $L$ is 
applied in a direction opposite to the direction of current 
in the ratchet. The overall potential is then  $V(x)=-[V_0(x) - x L]$. 
As long as the load is less than the stopping force $L_s$ current flows against
the load and the ratchet does work.  Beyond the stopping force 
the current flows in the same direction as the load and hence no 
useful work is done. Thus in the operating range of the load, 
$0<\,L<\,L_s$, the Brownian particles move in the direction 
opposite to the load and the ratchet does useful 
work~\cite{kamgawa-parrondo}. The average rate of work done over a 
period is given by~\cite{kamgawa} 
\begin{eqnarray}\label{Eout}
E_{out}= L [j^+ + j^-]\,.
\end{eqnarray}
The thermodynamic efficiency of energy transduction is 
~\cite{sekimoto,parrondo} 
\begin{equation}\label{eta}
\eta_{t}=  \frac{L [j^+ + j^-]}{F [(\frac{1+\epsilon}{1-\epsilon})
j^+-j^-]}. 
\end{equation}  
At very low temperatures or in the deterministic limit and also in 
the absence of applied load, the barriers in the forward 
direction disappears when $\frac{(1+\epsilon)}{(1-\epsilon)}F > 
\frac{Q}{\lambda_1}$ or $F > \frac{Q(1-\epsilon)}{\lambda_1(1+\epsilon)}$,  
and a finite current starts to flow in the forward direction. 
When $F > Q/\lambda_2$, the barriers in the backward direction 
also disappears and hence we now have a current in the backward 
direction as well leading to a decrease in the average current. 
In between the above two values of $F$, the current increases 
monotonically and peaks around $Q/\lambda_2$. In this range, 
a high efficiency is expected~\cite{sokolov,rkamj,pre,jstat}. In the limit 
when there is only forward current in the ratchet 
i.e. $j^+>>j^-$ and $L=0$ generalized efficiency reduces 
to Stoke's efficiency and is given by 
\begin{equation} \label{stokes}
\eta_{S}=\frac{(1-\epsilon)j^+}{(1+\epsilon)F}.
\end{equation}
In the present work we mainly focus on the case when the load $L=0$.

For the case of adiabatic rocking the ratchet can be considered 
as a rectifier~\cite{sokolov} and in the deterministic limit 
of operation and with zero applied load when $F$ is 
in the range $\frac{Q}{\lambda_2} > F > \frac{Q(1-\epsilon)}
{\lambda_1(1+\epsilon)}$, finite forward current alone exists 
and the analytic expression for current is given by
\begin{equation}\label{current}
j^{+}=\frac{1}{2}{\Big[{\frac{\lambda_1^2}
{(1+\epsilon)F\lambda_1-Q(1-\epsilon)}+
\frac{\lambda_2^2}{(1+\epsilon)F\lambda_2+Q(1-\epsilon)}}\Big]}^{-1} 
\end{equation}

Thus, Eqns.~\ref{stokes} and ~\ref{current} give an analytical expression 
for the Stoke's efficiency in the deterministic limit. We take all 
physical quantities in dimensionless units. The energies are 
scaled with respect to the height of the ratchet potential, $Q$; all lengths 
are scaled with respect to the period of the potential, $\lambda$, which is 
taken to be unity and we also set $\gamma=1$ . In the following section we 
present our results followed by discussions~\cite{pre,jstat,overdamp}.

\section{Results and Discussions}\label{sect3} 

\begin{figure}[hbp!]
 \begin{center}
\input{epsf}
\includegraphics [width=3in,height=2.5in,angle=270]{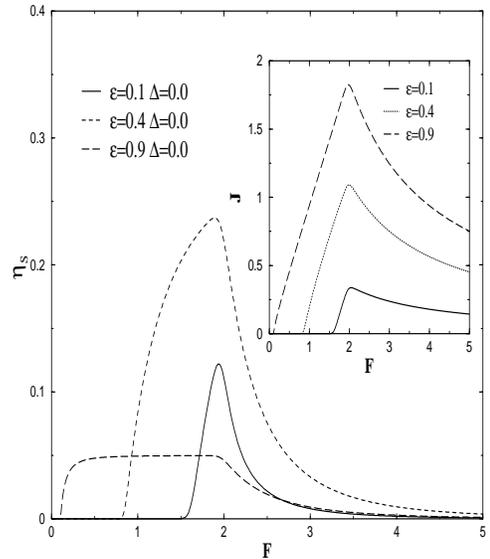}
\caption{Plot of efficiency as a function of $F$ at $T=0.01$ for
  different $\epsilon$ values with $\Delta=0.0$. Inset shows the behavior of
  current for the parameters given in the figure } 
\label{gen_FD0-inset}
 \end{center}
 \end{figure} 
\begin{figure}[htp!]
 \begin{center}
\input{epsf}
\includegraphics [height=2.6in,width=3.in,angle=270]{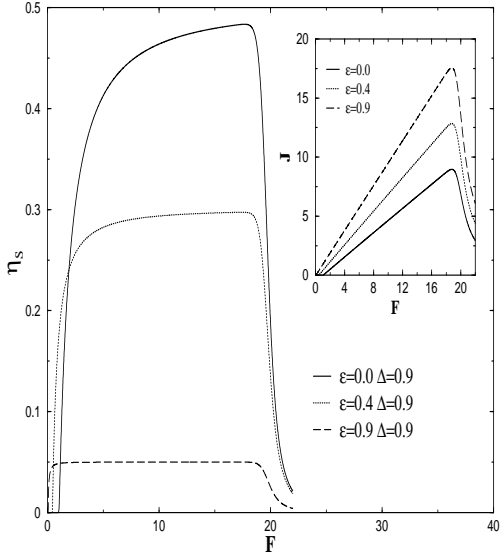}
\caption{Plot of efficiency as a function of $F$ at $T=0.01$ for
  different $\epsilon$ values $\Delta=0.9$. Inset shows the behavior of
  current for the parameters given in the figure. } 
\label{gen_FD9-inset}
 \end{center}
 \end{figure}

In Fig.~\ref{gen_FD0-inset} and Fig.\ref{gen_FD9-inset} 
we plot generalized efficiency in the absence of load or Stoke's 
efficiency as a function of $F$ 
for different values of $\epsilon$ at $T=0.01$ for symmetric $\Delta=0.0$ 
and asymmetric potential $\Delta=0.9$ respectively. 
As we increase $F$ in the interval from zero to 
$F_{min} = \frac{Q(1-\epsilon)}{\lambda_1(1+\epsilon)}$ the current 
is almost zero since barriers to motion exist in both 
forward (right) and backward (left) directions. This critical value 
of $F$ will decrease as we increase $\epsilon$ as seen from
Figs.~\ref{gen_FD0-inset} and ~\ref{gen_FD9-inset}. For $F>F_{min}$ barriers 
to the right disappears and as a consequence the current (inset) increases 
as a function of $F$ till  
$F_{max} = \frac{Q}{\lambda_2}$ beyond which the current also starts flowing 
in the backward direction. The behaviour of Stoke's efficiency 
reflects the nature of current (cf Eqn.~\ref{stokes}). Note 
that the value of $F_{max}$ does not depend on the time asymmetry parameter 
$\epsilon$, as is clear from Figs.~\ref{gen_FD0-inset} and 
~\ref{gen_FD9-inset}. Beyond $F_{max}$, barriers to motion in both 
the directions disappear and currents as well as generalized efficiency 
starts decreasing beyond $F_{max}$. We have seen that  input energy increases 
monotonically with $F$ for all the parameters. Hence the qualitative behaviour 
of current is reflected in the nature of generalized efficiency.
From the plot we see that the dependence of
generalized efficiency on $\epsilon$ is not in a chronological manner. High
$\epsilon$ value need not correspond to high generalized efficiency. 
For a given $\epsilon$ the current and Stoke's efficiency 
exhibits a peak around $F_{max}$.

We see from Eqn.~\ref{current} that for $\lambda_1>>\lambda_2$ 
 (i.e., large spatial asymmetry) and $F_{min}<F<F_{max}$, the analytical 
results from Eqn.~\ref{current} for the forward current fraction 
is simply given by $j_+ = \frac{(1+\epsilon)F}{2\lambda_1}$, 
while the Stoke's efficiency becomes $\eta_S =
\frac{1-\epsilon}{2\lambda_1}$. It is obvious from  
Fig.~\ref{gen_FD9-inset} that in this domain, $j+$ is a linear 
function of $F$ (inset) while $\eta_S$ exhibits a plateau in this regime.
This plateau regime is clearly observable for $\epsilon=0.9$ and
$\Delta=0.9$ as in Fig.\ref{gen_FD9-inset}. For these parameters, the ranges 
between $F_{min}$ and $F_{max}$ is large and moreover $\lambda_1>>\lambda_2$.
The value of $\eta_S$ at the plateau is $0.05$, which 
is consistent with the analytical result.
\begin{figure}[hbp!]
 \begin{center}
\input{epsf}
\includegraphics [width=3in,height=2.5in,angle=270]{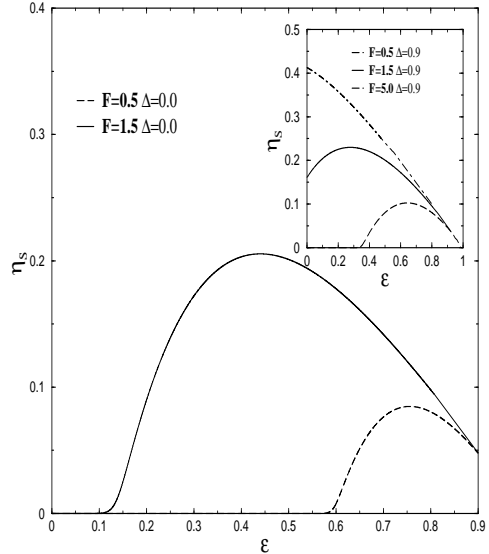}
\caption{Plot of Stoke's efficiency as a function of $\epsilon$ at 
$T=0.01$ for $F = 0.5$ and $1.5$ for $\Delta=0.0$. Inset shows the 
behaviour of efficiency for $F=0.5\,,\,1.5$ and $F=5.0$ but with $\Delta=0.9$
for the parameters given in the figure. } \label{gen-eps-inset}
 \end{center}
 \end{figure}
In contrast to the nature of $\eta_S$, we see that the average current, for a 
given $F$ and $\Delta$, always increases as $\epsilon$ is increased, see 
insets of Figs.~\ref{gen_FD0-inset} and \ref{gen_FD9-inset}. However, we 
see from Eqn.~\ref{stokes} that 
$\eta_{S}$ also depends on $\epsilon$ through the factor 
$\frac{(1-\epsilon)}{(1+\epsilon)}$ which is a decreasing function of
$\epsilon$ and hence the existence of optimal value of $\epsilon$ for 
$\eta_{S}$ is understandable. For a large spatial asymmetry,
$\Delta=0.9$, in the ratchet potential the magnitude of the 
average currents are quite large even for given $\epsilon$ 
as compared to the case when $\Delta$ is small.

From Fig.~\ref{gen_FD0-inset} we notice that the optimum value of 
generalized efficiency obtained is around $20\%$. This is the case of 
symmetric potential driven by temporally asymmetric force. From 
Fig.~\ref{gen_FD9-inset}, we notice that the inclusion of spatial 
asymmetry in the potential helps in enhancing the generalized efficiency 
and we can obtain an optimal value of nearly $50\%$ for 
efficiency in a particular parameter space.

In Fig.~\ref{gen-eps-inset} we plot the generalized 
efficiency with zero load (or Stoke's efficiency $\eta_S$) as a function of 
$\epsilon$ for different $F$ and symmetric potential. We have taken 
$T=0.01$ so as to be closer to the deterministic limit. 
The inset shows the same plot with asymmetric potential. 
We observe that for a given value of $F$,
only those $\epsilon$ values contribute to $\eta_S$ for which $F>F_{min}$. 
The minimum value of $\epsilon$ is given by 
$\epsilon_{min} = \frac{(Q-\lambda_1F)}{(Q+\lambda_1F)}$. For 
larger $F$, $\epsilon_{min}$ shifts to a smaller value as can be seen easily
from the figure. Moreover from Eqn.\ref{stokes} we can see that 
as $\epsilon$ approaches 1, the $\eta_S$ approaches 
zero (even though, strictly speaking, the 
$\epsilon \rightarrow 1$ limit is pathological). Thus, for the chosen 
parameter values the $\eta_S$ exhibits 
a peaking behavior. Note that the current vanishes 
due to the spatial symmetry of the potential in the limit 
$\epsilon \rightarrow 0$.

We now study the case when there is a spatial asymmetry which is shown
in the inset of Fig.~\ref{gen-eps-inset}. Here, a finite 
current can arise even when $\epsilon = 0$ provided force 
$F > \frac{Q}{\lambda_1}$. Thus $\eta_S$ in this regime 
can have finite value at $\epsilon = 0$ and can show a peaking 
behaviour. For $F >> \frac{Q}{\lambda_1}$, efficiency shows a 
monotonically decreasing behavior as a function of 
$\epsilon$. {\it{This clearly brings out the fact that in 
certain parameter ranges, time-asymmetric driving need not 
help in enhancing $\eta_S$ in the presence of spatially 
asymmetric potential}}. In the range $F < \frac{Q}{\lambda_1}$, 
currents are zero at $\epsilon=0$; thus $\eta_S$
exhibits a peaking behaviour with a value of 
zero for $\epsilon = 0$ and $\epsilon = 1$ in accordance with 
Eqn.~\ref{stokes}. These results show that $\eta_S$ is not a monotonically 
increasing function of $\epsilon$.
\begin{figure}[htp!]
\begin{center}
\input{epsf}
\includegraphics [width=2.8in,height=2.5in]{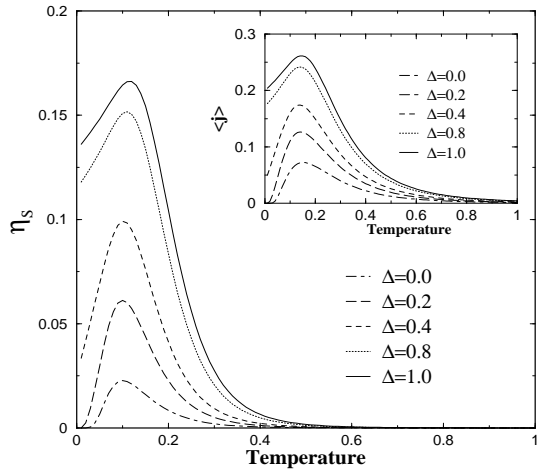}
\caption{Efficiency as a function of temperature for $F=1.0$ and
 $\epsilon=0.2$ for varying $\Delta$. Inset
 shows the behaviour of current with temperature for 
the same set of parameters.} 
\label{gen-TvaryD}
\end{center}
\end{figure}

We next discuss the behaviour of $\eta_S$ with temperature.
In Fig.~\ref{gen-TvaryD} we plot $\eta_S$
as a function of temperature for a fixed $F=0.1$ and temporal
asymmetry $\epsilon=0.9$ but with varying potential asymmetry $\Delta$. 
In most of the generic parameter space, we observe that temperature (or noise)
facilitates $\eta_S$ which quite is opposite to the generically 
observed behavior of \thmeff~\cite{jstat}. For example, if we take 
a particular curve say, $\Delta=0.0,F=0.1, \epsilon=0.9$,  
we can see that the value of efficiency is zero at $T=0$. 
This is because of the presence of the barriers in either directions during 
rocking. Thus when $F < F_{min}$, efficiency (current) is zero and as 
temperature is increased current starts to build up since 
Brownian particles can readily overcome the barriers to 
the right in the adiabatic limit. Beyond a certain $T$, 
current or efficiency will start to subside again as too 
much of noise will help the particle to overcome the barriers 
in both directions, thereby reducing the ratchet effect. Hence both 
current and generalized efficiency will fall. {\it{Thus for $F < F_{min}$ 
temperature always facilitates $\eta_S$}}.
\begin{figure}[htp!]
 \begin{center}
\input{epsf}
\includegraphics [width=3in,height=2.5in,angle=270]{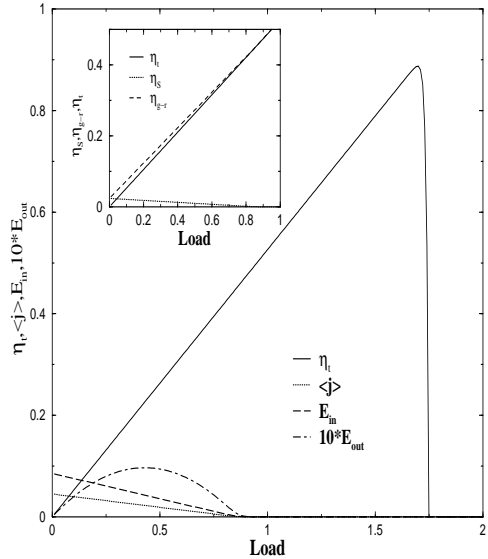}
\caption{Plot of thermodynamic efficiency, input energy, current and output
energy as a function of load for $\epsilon=0.9$, $F=0.1$, $\Delta=0.9$ and
$T=0.01$. The $E_{out}$ is scaled by a factor of $10$ to make it 
comparable to the scale chosen. Inset shows a comparison of 
the behaviour of thermodynamic, 
generalized and Stoke's efficiency for the same set of parameter
values. } 
\label{need}
 \end{center}
 \end{figure}
With increase in spatial asymmetry we see a finite current even 
when temperature is zero. This is because of the disappearance 
of the barriers to motion in the forward direction. However Stokes 
efficiency increases and shows a peaking behavior as a function of 
temperature even in this range. 
In Fig.~\ref{gen-TvaryD} note that peak value of current (inset) shifts to the 
right as we increase $\Delta$.  With increase in $\Delta$ 
barriers to the left increases and hence to overcome 
these larger barriers higher temperature is required. Only above these 
temperatures does current in backward direction begin to 
flow causing decrease of average current. Hence it is 
understandable that peak in average current shifts 
to higher $T$ with increase in $\Delta$.

When $F>F_{max}$, the barriers in both directions disappear. We have
separately verified that in this case both $\eta_S$ and the net current
decreases monotonically as a function of temperature.
 
In the end we discuss  briefly the differences between 
the nature of thermodynamic efficiency ($\eta_t$), 
generalized efficiency ($\eta_{g-r}$) and 
stokes efficiency ($\eta_{S}$). 
For the sake of comparison, we apply a load to the system.
In Fig.~\ref{need} we plot the $\eta_t$, net current $<j>$, input ($E_{in}$)
and output ($E_{out}$) power as a function of 
load for $\epsilon=0.9$, $\Delta=0.9$, 
$F=0.1$ and $T=0.01$. In the inset we plot the
$\eta_{g-r}$, $\eta_{S}$ and $\eta_t$ as a function of load 
for the same set of parameters so as to have a comparative 
idea of the behaviour of the different definitions of efficiency.

We notice that the thermodynamic efficiency increases with load from 
zero and exhibits a high value ($90\%$) just before the stopping force or
critical load, the range within which $\eta_t$ is defined. 
In contrast, $\eta_{S}$ (shown in the inset) has 
a finite value even when the load force is zero and then decreases 
monotonically with load. $\eta_{S}$ is almost zero and so is the 
velocity and current in the range where $\eta_t$ 
is very high. The magnitudes of current, $\eta_S$, $E_{in}$ 
and $E_{out}$ are very small near the stopping force, 
and hence are not observable on Fig.~\ref{need} due to the scale used. Both 
$\eta_{g-r}$ or $\eta_S$ starts with a finite value 
when load $L=0$ and it differs from $\eta_t$ in the low load limit. 
When the load value increases, $\eta_{g-r}$ also increases and 
at larger values of $L$ (near stopping force) it coincides with 
$\eta_t$. The main contribution to $\eta_{g-r}$ comes essentially from 
the work done against the load since velocity of particle 
is almost negligible and thus average power needed to move 
against the frictional drag becomes very small.
 
Another observation is that the average work done exhibits a peaking behavior 
where the thermodynamic efficiency is small and it is vanishingly 
small where the latter peaks. The average input power and current 
monotonically decreases with the load.
The figure clearly indicates that high thermodynamic 
efficiency does not lead to higher currents /  work / 
Stoke's efficiency. These results clearly bring out 
glaring differences between different 
definitions of efficiency as they are based on 
physically different criteria of 
motor performance ~\cite{derenyi,munakata,munakata1,machura2}.

\section{Conclusions}\label{sect4}

We have studied the generalized efficiency in an 
adiabatically rocked system in the presence of spatial and temporal 
asymmetry. The Stoke's efficiency exhibits a value 
of $50\%$ by fine tuning the parameters. Moreover, in 
a wide range of parameters this efficiency is much 
above the subpercentage regime. We have shown that 
in a wider parameter space temporal asymmetry may or
may not facilitate the generalized efficiency whereas 
generically, temperature facilitates it. 
In the regime of parameter space where the current is 
zero in the deterministic limit, temperature always facilitates 
Stoke's efficiency. In contrast, if the current is 
non-zero in the deterministic regime, depending on the 
parameters, it may happen that Stoke's efficiency monotonically 
decreases with temperature. 
The obtained high values for both the thermodynamic and generalized 
efficiency is attributed to the effect of suppression of current in the 
backward direction. Recently, it has also been shown that 
the same effect in these ratchet systems leads to enhanced 
coherency or reliability in transport.~\cite{unpub}. 
In conclusion, in suitable parameter ranges, 
our system exhibits high values for all the performance 
characteristics, namely, Stoke's efficiency, thermodynamic 
efficiency along with a pronounced transport coherency.

\section{Acknowledgement}

One of us (AMJ) thanks Dr. M. C. Mahato for useful discussions.

\end{document}